\documentclass[a4paper]{aa}
\usepackage{psfig}
\usepackage{epsfig}
\usepackage{graphicx}

\def \xmm {XMM--Newton}
\def \sax {BeppoSAX}
\def \src {EXO\thinspace0748--676}

\def \hcm {\hbox {\ifmmode $ atom cm$^{-2}\else atom cm$^{-2}$\fi}}
\def \arcmin {\hbox{$^\prime$}}

\def \chisq {$\chi ^{2}$}

\def\approxgt{\mathrel{\hbox{\rlap{\lower.55ex \hbox {$\sim$}}
        \kern-.3em \raise.4ex \hbox{$>$}}}}
\def\approxlt{\mathrel{\hbox{\rlap{\lower.55ex \hbox {$\sim$}}
        \kern-.3em \raise.4ex \hbox{$<$}}}}

\begin{document}

%\thesaurus{(02.01.2; 08.09.2; 08.14.1; 10.07.3; 13.25.2; 13.25.3)}

\title{The broad-band X--ray spectrum of the dipping Low Mass 
X-ray Binary \src}

\author{L. Sidoli\inst{1}
        \and A. N. Parmar\inst{2}
        \and T. Oosterbroek\inst{3}    
}
\offprints{L.Sidoli (sidoli@mi.iasf.cnr.it)}

\institute{
        Istituto di Astrofisica Spaziale e Fisica Cosmica --
        Sezione di Milano ``G. Occhialini" -- IASF/C.N.R. \\
         via Bassini 15, I-20133 Milano, Italy
\and
       Astrophysics Mission Division, Research and Scientific Support 
       Department of ESA, ESTEC, \\
       Postbus 299, NL-2200 AG Noordwijk, The Netherlands
\and
       INTEGRAL Science Operations Centre,
       Science Operations and Data Systems Division,
       Research and Scientific
       Support Department of ESA, ESTEC,
       Postbus 299, NL-2200 AG Noordwijk, The Netherlands}

\date{Received 7 January 2004 ; Accepted 13 August 2004 }

\authorrunning{L. Sidoli et al.}

\titlerunning{{The broad-band spectrum of \src}}

\abstract{We present results of a 0.1-100~keV \sax\ observation of the
dipping low-mass X-ray binary (LMXRB) \src\ performed in 2000 November. 
During the observation \src\ exhibited X-ray eclipses, type I X-ray bursts and
dipping activity over a wide range of orbital phases. 
The 0.1--100~keV  ``dip--free''(ie. dipping and eclipsing intervals excluded) 
spectrum is complex,
especially at low-energies where a soft excess is present. Two 
very different spectral models give satisfactory fits. The first
is the progressive covering model, consisting of separately absorbed
black body and cut-off power-law components. This model is 
often used to study the dipping activity in LMXRB. 
The second model is an absorbed cut-off power-law together with a moderately ionized
absorber with a sub-solar abundance of Fe and a 2.13~keV absorption feature 
(tentatively identified with Si\,{\sc xiii}). This ionized absorber may
be the same feature as seen by {\it Chandra} during dips from \src. 
The fact that these two physically very different models both give acceptable
fits to the data and plausible values of the best-fit parameters,
indicates how difficult it is to reliably model such complex 
dipping spectra with moderate spectral resolution data. 
The blackbody component required by the progressive covering 
LMXRB model probably has a more complex underlying nature, 
due to the strong effects of ionized absorption present.
}
\maketitle

\section{Introduction}
\label{sect:intro}

\src\ is a transient low-mass X-ray binary (LMXRB) discovered with
EXOSAT (Parmar et al. \cite{p:86}). 
It shows X-ray eclipses every 3.82 hr, complex dipping activity
and type I X-ray bursts, from which a distance of 10 kpc has been estimated 
(Parmar et al. \cite{p:86}; Gottwald et al. \cite{g:86}). 
Quasi-periodic oscillations (QPOs) at Hz and kHz frequencies were
discovered using R-XTE (Homan et al. \cite{h:99}; Homan \& van der Klis \cite{h:00}).
A summed \xmm\ Reflection Grating Spectrometer (RGS) spectrum
of 28 type I bursts from \src\ revealed absorption features 
identified with Fe~{\sc xxv} and {\sc xxvi} and O~{\sc viii} 
transitions all with a gravitational red-shift of 0.35 (Cottam et al. \cite{cpm:02}).
For the astrophysically plausible range of neutron star masses the
measured red-shift is completely consistent with models of neutron
stars composed of normal neutral matter, while it excludes some models
in which neutron stars are made of more exotic matter.

The spectral and temporal complexity of the soft X-ray spectrum of \src\ 
was already hinted at in early ROSAT and ASCA observations
(Schulz \cite{s:99}; Thomas et al. \cite{t:97}; 
Church et al. \cite{c:98}). 
The vastly improved narrow line sensitivity of the RGS has allowed
the 0.3--2.5~keV X-ray spectrum of \src\ to be studies in detail
revealing the presence of
both absorption and emission features from ionized Ne, O and N. The
line producing material is
probably located well away from the orbital plane (Cottam et al. \cite{c:01}).
The lines are broadened ($\sim$1000 km~s$^{-1}$) and exhibit a direct 
correlation between the velocity broadening and the degree of ionization. 
The velocity shift of the emitting plasma is $<$300~km~s$^{-1}$.
The RGS lightcurve does not show any orbital modulation, indicating
that the source of the soft continuum is extended.
0.3--10~keV observations obtained with the \xmm\ European Photon Imaging Camera
(EPIC) revealed a varying lightcurve with soft X-ray flares together 
with dipping activity, X-ray eclipses
and type I X-ray bursts (Bonnet-Bidaud et al. \cite{bb:01}).
The  ``persistent" (ie. dipping and eclipsing intervals excluded)
spectrum was interpreted using a two-component model
consisting of an extended thermal ($kT$ = 0.64 keV) component with a radius of 
$3\times 10^{10}$~cm and a power-law with a 
photon index, $\Gamma$, of 1.35. The power-law is affected by additional
absorption, $N_{\rm H}$, of $6.4\times10^{22}$~cm$^{-2}$ and is probably
produced in a more compact Comptonizing accretion disk corona with a 
radius of $2\times10^{8}$~cm.
The thermal component may have strongly non-solar abundances 
of N (10$\pm{2}$), O (0.6$\pm{0.1}$),
Ne (0.02$\pm{0.02}$) and  Mg (0.31$\pm{0.23}$).

A   $Chandra$ observation of \src\ with the High-Energy Transmission Grating 
Spectrometer (HETGS) revealed several strong discrete spectral signatures 
of a photo-ionized plasma in orbit around the neutron star
(Jimenez-Garate et al. \cite{jg:03}). 
Recombination lines from H-like and He-like
O, Ne and Mg were detected, with a mean velocity broadening of 
$\sim$750~km~s$^{-1}$. The continuum requires compact soft and hard sources
together with an extended recombination region associated with the 
thickened outer
regions of the accretion disk where the accretion stream impacts. 
The spectral changes during dips require the presence of both a neutral
and an ionized absorber (Jimenez-Garate et al. \cite{jg:03}). 

\section{Observations and Data Reduction}
\label{sect:obs}

\src\ was observed with \sax\
between 2000 November 4 16:52 and November 6 07:44 UTC, 
with an on-source time of 66~ks.
We present results 
from the Low-Energy Concentrator Spectrometer (LECS;
0.1--10~keV; Parmar et al. \cite{p:97}), the Medium-Energy Concentrator
Spectrometer (MECS; 1.8--10~keV; Boella et al. \cite{b:97}),
High Pressure Gas Scintillation Proportional Counter
(HPGSPC; 5--120~keV; Manzo et al. \cite{m:97}) 
and the Phoswich Detection System (PDS; 15--300~keV;
Frontera et al. \cite{f:97}) instruments.
All these instruments are coaligned and collectively referred
to as the Narrow Field Instruments.
The MECS and the LECS are grazing incidence 
telescopes with imaging gas scintillation proportional counters in
their focal planes. 
The non-imaging HPGSPC consists of a single unit with a collimator
that was alternatively rocked on- and 180\arcmin\ off-source 
every 96~s during most of the observations. During the observations
in the year 2000 the collimator remained on-source.
The 
PDS consists of four independent non-imaging 
units arranged in pairs each having a
separate collimator. Each collimator was alternatively
rocked on- and 210\arcmin\ off-source every 96~s during 
the observations.
The data were reprocessed using the SAX Data Analysis System (SAXDAS) 
version 2.0.0.
For the
spectral analysis counts have been extracted from circular regions with standard radii
(4$'$ for the  MECS and 8$'$ for the LECS).
The LECS and MECS spectra were rebinned to oversample the full
width half maximum of the energy resolution by
a factor 3 and to have additionally a minimum of 20 counts 
per bin to allow use of   $\chi^2$ statistics. 
The HPGSPC and PDS 
spectra were rebinned using the standard techniques in SAXDAS.
Response matrices appropriate for the sizes of the extraction regions 
were used.

%-------------------------------------------------------------------------------------------------
\begin{figure}[!ht]
\centerline{\psfig{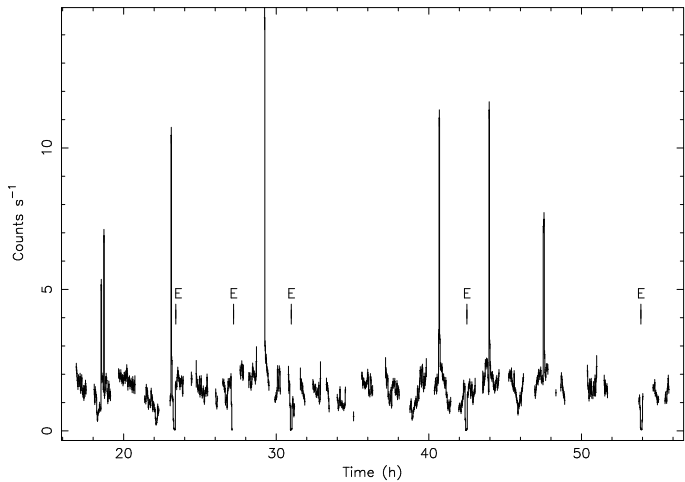}}
\vskip 0.0truecm
\caption{\src\ MECS 1.8--10~keV lightcurve with a binning of 128~s. 
A double burst and 5 single
bursts are clearly visible together with eclipses every 3.8~hr (marked by the letter ``E").}  
\label{fig:lc}
\end{figure}
%-------------------------------------------------------------------------------------------------

\section{Results}

%------------------
\begin{figure*}[!ht]
\centerline{\psfig{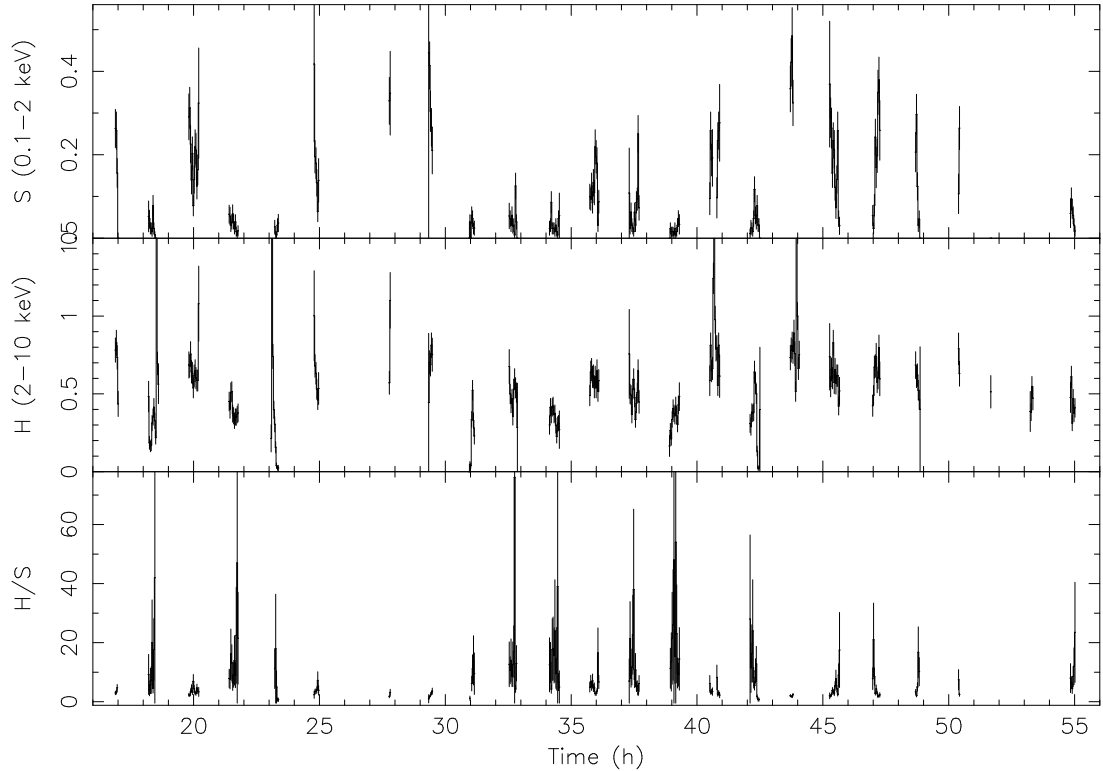}}
\vskip 0.0truecm
\caption{\src\ LECS 0.1--2 keV (``S'', upper panel) and
LECS 2--10~keV (``H'', middle panel) lightcurves, 
with bursting intervals excluded. 
The lower panel shows the hardness ratio, H/S.
The binning time is  128~s.}
\label{fig:lc2}
\end{figure*}

\begin{figure}[!ht]
\centerline{\psfig{figure=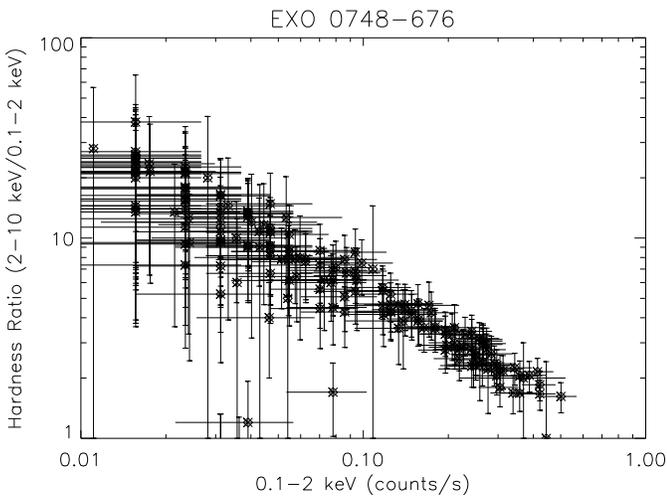,height=6.8cm,angle=0}}
\vskip 0.0truecm
\caption{Hardness ratio (2--10 keV LECS counts divided by those between
0.1--2 keV) versus the LECS 0.1--2~keV intensity.
Above a 0.1--2~keV count rate of around 
0.2~s$^{-1}$ the hardness ratio does not vary appreciably with count rate
}
\label{fig:hr}
\end{figure}

\src\ displays many kinds of variability such as X-ray eclipses, dips and
type I X-ray bursts (Fig.~1). In order to 
study the \src\ spectrum without the effects of this variability
it is necessary to identify dip-free intervals (the bursts and eclipses
are easy to detect).
Homan et al.~(\cite{h:03}) show that \src\ can display
low-level dipping activity which is not easily visible above
2~keV, but which can affect the overall spectrum. With this result in mind,
lightcurves in two different energy ranges were produced
(Fig.~\ref{fig:lc2}) with the bursting intervals excluded. 
Examination of the 0.1--2 keV LECS lightcurve (Fig.~\ref{fig:lc2},
upper panel) 
reveals strong variability throughout the observation. This
emphasizes the difficulty in identifying dip--free intervals
using only a visual inspection of the source lightcurves. 
In order to investigate whether variations in the source 
hardness ratio provide a
reliable means of identifying dip-free intervals,
the LECS 2--10~keV counts were divided by those
between 0.1--2~keV to create hardness ratios which were 
plotted against the 0.1--2.0~keV count rate (Fig.~\ref{fig:hr}).
The dipping intervals are clearly evident 
due to their higher hardness ratios. 
For count rates $\approxgt$0.2~s$^{-1}$ the hardness ratio does not
vary appreciably. Thus, by selecting events 
when the count rate was above this intensity threshold, it is
possible to obtain the dip--free \src\ spectrum.
Additional confidence that this selection corresponds to the 
``dip--free" spectrum is obtained from considering
spectra extracted from intervals corresponding to nearby 
intensity selections (e.g., 0.2--0.3 count s$^{-1}$,
0.3--0.4 count s$^{-1}$, and $>$0.4 count s$^{-1}$). These spectra 
do not show significant differences
from that extracted from intervals with $>$0.2~count~s$^{-1}$.

%-------------------------------------------------------------------------------------------------
%-------------------------------------------------------------------------------------------------

Thus, the extraction of an eclipse and dip free
spectrum was performed using
only time intervals where the source displays 
a constantly low hardness ratio,
corresponding to intervals
when the LECS 0.1--2~keV count rate was $>$0.2~s$^{-1}$ when 
rebinned on a timescale of 128~s. Since the LECS was
only operated during spacecraft night when the other 3 instruments were
also operated, the same time intervals were then used to extract 
corresponding MECS, HPGSPC and PDS
spectra, leading to exposure times of 4.3~ks for all 4 instruments.

The data were then selected in the energy ranges
0.1--8.0~keV (LECS), 1.8--10~keV (MECS), 8--20 keV (HPGSPC)
and 15--100~keV (PDS) and rebinned using standard procedures.
The resulting  background-subtracted count rates were
0.9, 1.6, 2.3 and 3.4~$s^{-1}$
for the LECS, MECS, HPGSPC and PDS, respectively.
In the spectral fitting, factors were included
to account  for normalization
uncertainties between the instruments.
These factors were constrained
to be within their usual ranges during the fitting.
All spectral uncertainties
and upper-limits are given at 90\% confidence.
Spectral analysis was performed with {\sc xspec} v.11
software package. 
 
\src\ is detected in the PDS up to an energy of 100~keV.
An absorbed power-law model does not account for the broad-band emission,
leaving large positive residuals below 1 keV and around 20~keV, 
together with negative
residuals $\approxgt$30 keV, indicative of a high-energy cut-off 
(\chisq = 892  for 176 degrees of freedom, d.o.f.).
Adopting a cut-off power-law model
resulted in a significantly better fit (\chisq/d.o.f. = 407/175), although the
fit is still
unacceptable because of the presence of a large soft excess below 
1~keV.
We note that a satisfactory fit (\chisq/d.o.f. = 160.1/154) 
to the 2--100~keV spectrum can
be obtained with an absorbed 
cut-off power-law model
with $N_{\rm H}$ = $2\times10^{22}$~cm$^{-2}$,  
$\Gamma$=1.3 and a cut-off energy, $E_{\rm c}$, of 44~keV. 
 
The 0.1--100~keV spectrum
was then modeled using different types of two-component models,
always including the cut-off power-law to account for the high energy part
of the spectrum. The soft excess was modeled using black-body
and multicolor disk-blackbody (Mitsuda et al. \cite{m:84}) components, 
an ionized absorber and partial covering absorption.
However, unacceptable fits (\chisq$_{red}$$>$1.4) were always obtained. 
The \xmm\ observations reported in 
Bonnet-Bidaud et al.~(\cite{bb:01}) revealed a complex persistent
spectrum, deconvolved with a two-component model consisting of
a Raymond-Smith (RS) thermal emission ($kT_{\rm RS}$ = 0.64 keV) 
and a power-law component ($\Gamma$ = 1.35). The two components were 
differently absorbed with $N_{\rm H}$ = 6.4$\times$10$^{22}$~cm$^{-2}$ 
for the 
power-law, and $N_{\rm H}$ = 0.15$\times$10$^{22}$~cm$^{-2}$ for the
thermal plasma. This model was fit to our data, with the same parameters
(but with the normalizations free), but it does not well
account for the \sax\ data.
A {\sc vmekal} model (e.g., a thermal plasma model with 
variable metal abundances) was tried next, since
Bonnet-Bidaud et al. (\cite{bb:01}) require non-solar 
abundances of N (10$\pm{2}$), O (0.6$\pm{0.1}$),
Ne (0.02$\pm{0.02}$) and  Mg (0.31$\pm{0.23}$) in their fits. 
The abundances of N, O, Ne
and Mg were allowed to vary during the fitting, but could not be
usefully constrained, nor a significantly better fit obtained. 
Large positive residuals below 1~keV still are present.

\begin{figure*}[!ht]
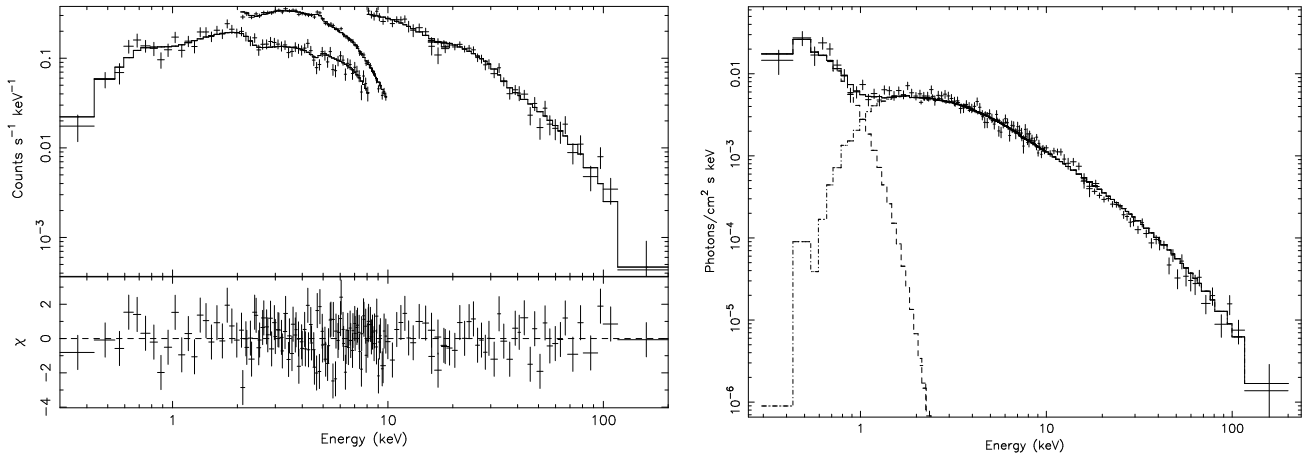

\hbox{\hspace{.0cm}
\includegraphics[height=8.8cm,angle=-90]{lsidoli_f5a.ps}
\hspace{0.2cm}
\includegraphics[height=8.0cm,angle=-90]{lsidoli_f5b.ps}}
\caption[]{0.1--100~keV \sax\ spectrum of \src\ fitted
with the progressive covering model proposed by Church et al. (1998)
(see Table~\ref{tab:spec_church} for the results).
The left panels show the best-fit count spectrum and the
residuals (in units of standard deviation). The right panel shows the
photon spectrum. 
}
\label{fig:spec_church}
\end{figure*}

Another model successfully used in the modeling of \src\ spectra
(and other dipping LMXRBs) was introduced by Church et al. (\cite{c:98})
and consists of a blackbody and an extended Comptonizing corona
described by a power-law spectrum (progressively covered by absorbing matter
during dipping activity). Applying this model to the 
0.1--100~keV ``dip--free" emission
resulted in a good description of the spectrum (\chisq/d.o.f. = 185.7/170).
Contrary to the ``standard" Church et al.~(\cite{c:98}) model, where the 
``dip--free" emission is un-covered,
extra-absorption of the Comptonized component (see Table~1 and 
Fig.~\ref{fig:spec_church}) is required
to obtain a satisfactory fit to the data.

\begin{figure*}[!ht]
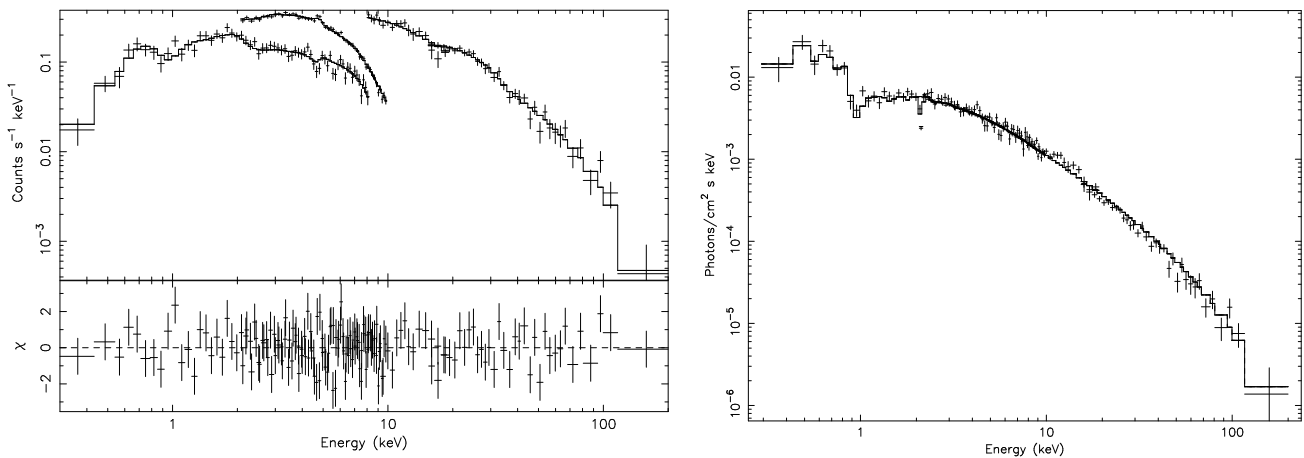

\hbox{\hspace{.0cm}
\includegraphics[height=8.8cm,angle=-90]{lsidoli_f6a.ps}
\hspace{0.2cm}
\includegraphics[height=8.0cm,angle=-90]{lsidoli_f6b.ps}}
\caption[]{0.1--100~keV \sax\ spectrum of \src\ fitted
with a cut-off power-law and an ionized absorber
(see Table~\ref{tab:spec_ionized} for the parameters).
The left panels show the best-fit count spectrum and the
residuals (in units of standard deviation). The right panel shows the
photon spectrum.
}
\label{fig:spec_ionized}
\end{figure*}

%%%%%%%%%%%%%%%%%%%%% ---- IONIZED ABSORBERS
Finally, we tested the possibility that the soft-excess is due 
to the presence of an ionized absorber. 
Adopting the {\sc absori} model in  {\sc xspec} 
(Done et al. \cite{d:92}) 
and fixing the ionizing power-law photon index to that 
of the cut-off power-law component, 
we obtained a satisfactory fit (\chisq/d.o.f. = 203.2/173) 
to the data with the following parameters: 
interstellar column density $N_{\rm H}$ = $10^{21}$~cm$^{-2}$,
an intrinsic absorber with a column density 
$N_{\rm absori}$ = (5$\pm{1})$$\times$$10^{22}$~cm$^{-2}$, an absorber ionization
state $\xi$ = 350$\pm{90}$~erg~cm~s$^{-1}$  ($\xi$ = L/nR$^{2}$),
a cut-off power-law with $\Gamma$ = 1.3$\pm{0.1}$  and
$E_{\rm c}$ = $45\pm{10}$ keV.
Since there are still strong negative residuals around 2.1--2.2~keV, 
%(see Fig.~\ref{fig:noline}), 
we included a narrow negative Gaussian in the model.
This led to a significantly better fit (\chisq/d.o.f. = 184.4/170), with 
{\sc absori} and cut-off power-law parameters substantially unchanged, 
plus an absorption
feature with a centroid at 2.13~keV and an equivalent width, 
$EW$, of $\sim$50 eV, which can be tentatively identified
with absorption from highly ionized Si. We note however that
this energy is close to the Au l-edges of the BeppoSAX mirrors and we cannot
exclude that this feature is due to incorrect instrumental modeling. 
A better fit can be found
if the Fe abundance in the ionized absorber 
is allowed to vary,
obtaining a sub-solar abundance of Fe (see Table~\ref{tab:spec_ionized} and
Fig.~\ref{fig:spec_ionized}).

%--------------------------------------------------------
\begin{table}
\begin{center}
\caption[]{Parameters for the 0.1--100~keV BeppoSAX ``dip--free" emission
spectrum of \src\ fitted with the Church et al.~(\cite{c:98}) 
model. $N_{\rm H}$ is the interstellar
column density,
$N_{\rm H}$(BB) and $N_{\rm H}$(POW)
are the column densities of the blackbody and power-law components,
$f$ is the covering fraction,
$\Gamma$ is the cut-off power-law photon index, 
and $E_{\rm c}$ is the high energy cut-off. 
Fluxes are uncorrected for absorption. 
The luminosity  
is {\it only} corrected for interstellar absorption and is for a 
distance of 10~kpc.
}
\begin{tabular}{ll}
\hline
\noalign {\smallskip}
Parameter & Value   \\
\hline
\noalign {\smallskip}
$N_{\rm H}$ (BB) $(10^{22}$ cm$^{-2}$)  		&  $0.12  \pm{0.01}$      \\
kT$_{\rm bb}$ (keV)                              & $ 0.13 ^{+0.02} _{-0.01}$  \\
R$_{\rm bb}$ (km)                   			&  $<$100   \\
$N_{\rm H}$ $(10^{22}$ cm$^{-2}$)  		&  $0.6  \pm{0.3}$     \\
$N_{\rm H}$ (POW) $(10^{22}$ cm$^{-2}$)  	&  $5.4  ^{+1.9} _{-1.7} $     \\
$f$ 						&  $0.4  ^{+0.2} _{-0.1} $     \\
$\Gamma$                      			 &   $1.34 ^{+0.12} _{-0.07}$      \\
$E_{\rm c}$          (keV)             		 &    $50 \pm{10}$    \\
Flux (2--10 keV) (${\rm erg~cm^{-2}~s^{-1}})$      		 &  $ 1.7\times 10^{-10}$    \\
Flux (0.1--100 keV) (${\rm erg~cm^{-2}~s^{-1}})$       		&  $  6.8 \times 10^{-10}$    \\
Luminosity (2--10 keV) (${\rm erg~s^{-1}})$          		&   $ 1.7\times 10^{36}$    \\
Luminosity (0.1--100 keV) (${\rm erg~s^{-1}})$          		&   $ 7.7\times 10^{36}$    \\
$\chi ^2$/d.o.f.                            			&  185.7/170           \\
\noalign {\smallskip}
\hline
\label{tab:spec_church}
\end{tabular}
\end{center}
\end{table}
%----------------------------------------------------------------

%% 
%--------------------------------------------------------
\begin{table}
\begin{center}
\caption[]{Best-fit parameters for the 0.1--100~keV BeppoSAX
``dip--free" spectrum of \src\ fitted with a cut-off power-law,
absorbed by an ionized absorber with a non-solar abundance of Fe. 
$N_{\rm H}$ is the interstellar column density,  $N_{\rm absori}$
the local absorbing column density due to the ionized
absorber where $\xi$ is the ionization parameter ($\xi$ = $L/nR^{2}$, where
$L$ is the luminosity of the illuminating source, $n$ is the absorbing 
plasma density and $R$ is the distance of the absorbing material from the 
ionizing source), $\Gamma$ is the cut-off power-law photon index and
$E_{\rm c}$ is the high energy cut-off. $E_{\rm line}$, $\sigma$ and 
$I_{\rm line}$ are the absorption line energy, width and intensity,
respectively.
Fluxes are uncorrected for absorption. 
The luminosity  
has been {\it only} corrected for interstellar absorption and is for a 
distance of 10~kpc.
}
\begin{tabular}{ll}
\hline
\noalign {\smallskip}
Parameter & Value   \\
\hline
\noalign {\smallskip}
$N_{\rm H}$ $(10^{22}$ cm$^{-2}$)  		&  $0.11  ^{+0.04} _{-0.04} $     \\
$N_{\rm absori}$ $(10^{22}$ cm$^{-2}$) 		 &  $8  ^{+5} _{-3} $     \\
$\xi$                                  		 & $ 450 ^{+170} _{-140} $   \\
Fe abundance                   			&  $0.23 ^{+0.30} _{-0.21}$   \\
$\Gamma$                      			 &   $1.35 ^{+0.10} _{-0.10}$      \\
$E_{\rm c}$          (keV)             		 &    $50 \pm{10}$    \\
$E_{\rm line}$   (keV)    			 &    $ 2.13 \pm{0.06}$    \\
$\sigma$ 	(eV)				&    $<$200     \\
$I_{\rm line}$  (${\rm  10^{-4} photons~cm^{-2}~s^{-1}})$   &    $-6 ^{+2} _{-6}$    \\
$EW$ (eV)                                          &    $ -50 ^{+10} _{-70}$    \\
Flux (2--10 keV) (${\rm erg~cm^{-2}~s^{-1}})$      		 &  $ 1.7\times 10^{-10}$    \\
Flux (0.1--100 keV) (${\rm erg~cm^{-2}~s^{-1}})$       		&  $  6.8 \times 10^{-10}$    \\
Luminosity (2--10 keV) (${\rm erg~s^{-1}})$          		&   $ 1.7\times 10^{36}$    \\
Luminosity (0.1--100 keV) (${\rm erg~s^{-1}})$          		&   $ 7.2\times 10^{36}$    \\
$\chi ^2$/d.o.f.                            			&  175.2/169            \\
\noalign {\smallskip}
\hline
\label{tab:spec_ionized}
\end{tabular}
\end{center}
\end{table}
%----------------------------------------------------------------

%----------------------FIG: church model SPECTRUM-------------------------------------
%-------------------------------------------------------------------------------

%--------FIG: ionized abs. model SPECTRUM--------------------------------
%-------------------------------------------------------------------------

\section{Discussion}
\label{sect:discussion}
 
We have performed for the first time broad-band spectroscopy
of the dipping, eclipsing, LMXRB \src.
The X-ray spectrum extends to at least 100~keV and exhibits a
cut-off at $50 \pm 10$~keV. At low-energies, the spectrum is 
particularly complex with a soft-excess that cannot
be easily fit. Various combinations of simple models were tried
without success. However, 
we did find that two models can adequately fit the data. In both
cases the high-energy component is modeled by a cut-off power-law, and
the difference is primarily how the low-energy part of the spectrum 
is modeled.
In the first case, following Church et al. (1998) this is as an
absorbed blackbody and additional partial covering of the cut-off power-law 
is required. Contrary to the Church et al.~(1998) model for
dipping LMXRB, additional absorption on the cut-off power-law
component is required. 
In the second model, an additional additive component
is not required and instead the cut-off power-law is absorbed by
moderately ionized material with a sub-solar Fe abundance. A 2.13~keV
absorption feature, which may be from highly ionized Si (Si\,{\sc xiii}),
or due to an inadequate modeling of the instrumental response, is also
required.
We note that the interstellar absorption determined with this model is 
consistent
with the galactic value measured in this direction 
(N$_{\rm gal}$ = 1.1$\times10^{21}$ cm$^{-2}$; 
Dickey \& Lockman \cite{dl:90}).
The fact that these two physically very different models both give acceptable
fits to the data and provide plausible values of the best-fit parameters
gives an indication of the complexity of the 
$\approxlt$2~keV \src\ spectrum and
the difficulties of modeling this region using data 
obtained from moderate spectral resolution instruments. 
In addition, these result indicate that the blackbody component 
required by the
Church et al.~(1998) LMXRB model probably has a more complex 
underlying nature, 
due to the strong effects of ionized absorption present in this source.

The normalization of the high energy cut-off power-law observed 
from \src\ during the \sax\ observation is about 1.4 times 
higher than that during the \xmm\ observation, performed about 7 
months earlier (Bonnet-Bidaud et al. \cite{bb:01}). 
The slopes of the high-energy part of the spectrum in these two observations 
are compatible. The \xmm\ observation demonstrated for the first time 
the existence 
of an extended soft thermal component from \src. 
We tried to fit the \sax\ observation with the \xmm\ best-fit, but were unsuccessful.
Indeed, the \sax\ 0.1--2~keV spectrum cannot be adequately fit with 
a hot thermal plasma model, even allowing a different absorption
with respect to the high-energy part (as in the \xmm\ observation). 
This probably means that during the \sax\ observation the extended thermal
halo was suppressed, while the Componization component  
became 1.4 times brighter.
The observed 0.5--10~keV (uncorrected for absorption) 
flux during the \sax\ observation 
($1.9\times10^{-10}$ erg~cm$^{-2}$~s$^{-1}$)
is about twice that observed during the \xmm\ observation 
($1.1\times10^{-10}$ erg~cm$^{-2}$~s$^{-1}$). 

If a thermal component similar to that observed in \xmm\ spectrum were 
present 
in \src\ during the
\sax\ observation (with the same temperature of 0.6 keV), it should have had 
an additional absorption of 1.1$\times10^{22}$ cm$^{-2}$ or
an emission measure, $A_{\rm RS}$, $<$3.3$\times10^{-3}$ (where $A_{\rm RS}$
is in units of  10$^{-14}$/(4$\pi$d$^2$) $\int$n$_{\rm e}$n$_{\rm H}dV$, 
where $d$ is the distance to the source (cm), $n_{\rm e}$  is the electron
density, and  $n_{\rm H}$ is the hydrogen density).
While the low abundance of Fe is consistent with the lack of any Fe line
detection by (Bonnet-Bidaud et al.~\cite{bb:01} with the \xmm\
RGS, the presence of ionized Si is not, since no emission 
lines from
Si were present in the RGS spectrum, 
leading the authors to conclude a very low abundance of this 
element.
This difference could be due to a different ionization status of the 
absorbing 
matter, or as mentioned earlier, may be an instrumental artifact due to
inadequate modeling of instrumental Au edges.
The ionized absorber detected in the ``dip--free" spectrum is probably 
the same 
as observed by $Chandra$ (Jimenez-Garate et al. \cite{jg:03}) during dips, 
and interpreted by them as a thickened photo-ionized region located i
n the outer 
accretion disk. Our results indicate that the effects of this ionized absorber
are also visible in the  ``dip--free" spectrum.

Using the best-fit value of the ionization parameter of the absorbing matter,
$\xi$$\sim$450~erg~cm~s$^{-1}$ ($\xi=L/nR^{2}$, where
$L$ is the luminosity of the illuminating source, $n$ is the absorbing 
plasma density and $R$ is the distance of the absorbing matter from the 
ionizing source), and the 2--10 keV luminosity 
($2.0\times 10^{36}$~erg~s$^{-1}$),
we obtain $nR^{2}$ = $\sim$4$\times10^{33}$~cm$^{-1}$.
Assuming that the absorbing plasma has the same dimensions
as the corona ($2\times10^{8}$~cm radius; Bonnet-Bidaud et al. \cite{bb:01}) 
we obtain a density
of $\sim$10$^{17}$~cm$^{-3}$ for the absorbing ionized plasma.
Alternatively, if we assume that this absorber is located in the outer 
accretion disk (as found
with $Chandra$, Jimenez-Garate et al. \cite{jg:03}, R$\sim$10$^{10}$~cm), 
we find a density
of $\sim$4$\times10^{13}$~cm$^{-3}$.

It is useful to compare our results with those of  
four \xmm\ observations of \src\ obtained in 2000--2001 recently reported 
in Homan et al.~(\cite{h:03}).
Due to the brightness of the source, serious pile-up problems occurred 
during the \xmm\ observations both
in the MOS and PN cameras,
limiting the spectral analysis to only one \xmm\ pointing
operated with the EPIC PN camera in small window mode (the same
observation was already analysed by Bonnet-Bidaud et al. (\cite{bb:01})).
Their analysis mostly concentrated on the spectral evolution during dips,
showing that current models (Church et al. \cite{c:98} and 
Bonnet-Bidaud et al. \cite{bb:01} models) 
are not able to account for the spectral changes
during dips.
We found similar difficulties in modeling the source emission even during
``dip--free" intervals. This suggests that other spectral components 
are probably
present especially at soft energies, and leading us to search for 
a new and different
spectral model to describe the source emission. 
In the \xmm\ observations analysed by Homan et al.~(\cite{h:03}) 
clear ``flaring" activity is present as demonstrated by the very 
high dynamical range in the 0.3--2~keV lightcurves. 
The complexity of our \sax\ ``dip--free" spectrum at soft energy 
is apparently 
unrelated with possible spectral mixing with flaring activity.
Indeed, comparing the source variability seen with \sax\ (a factor of less
than 2 compared to a factor $\sim$4 in the \xmm\ data) 
suggests that
any flaring activity during \sax\ observation was less pronounced.

\begin{acknowledgements}
%The \sax\ satellite is a joint Italian-Dutch programme. 
We would like to thank the anonymous referee for valuable comments and 
suggestions.
\end{acknowledgements}

\end{document}